# Finite temperature QCD: progress and outstanding problems

Carleton DeTar

Department of Physics, University of Utah
Salt Lake City, UT 84112, USA

I review recent progress in numerical simulations of finite temperature quantum chromodynamics and discuss the status of some outstanding problems. Included is (1) a discussion of recent results determining the temperature of the "phase transition" in full QCD, (2) a scaling analysis of the Polyakov loop variable, leading to the determination of a constituent quark free energy, (3) studies of critical behavior near the phase transition in two-flavor QCD, (4) a discussion of problems and new results in thermodynamic simulations with Wilson fermions, (5) recent results in pure gauge theory with a mixed fundamental/adjoint action, and (6) the nonperturbative determination of the equation of state with dynamical fermions included. Finally I mention briefly new developments in efforts to construct a phenomenology of deconfinement and chiral symmetry restoration, namely (7) the dual superconducting model and (8) the instanton model.

## 1. INTRODUCTION

This year has brought new, preliminary results from thermodyamic simulations with two quark flavors in the staggered fermion scheme at $N_t = 12$, new insights into and questions about the critical behavior at the phase transition with two quark flavors, a new nonperturbative determination of the equation of state for both pure Yang-Mills theory and QCD with two staggered fermions, and new results for Wilson thermodynamics with three quark flavors and at smaller lattice spacing, to mention a few highlights. Because companion reviews in this volume deal with Wilson thermodynamics[1] and the equation of state in pure Yang-Mills theory[2], I will treat these important topics only briefly. With staggered fermion thermodynamic data available now over a wide range of $N_t$ values it is tempting to try to systematize them through scaling relations. I offer an attempt toward this end.

## 2. TEMPERATURE OF THE PHASE TRANSITION

Of particular phenomenological interest is the temperature of the phase transition or crossover from the low to the high temperature regimes in QCD. With two flavors of light quarks in the stag-gered fermion scheme, there appears to be a dramatic crossover, but so far no evidence for a genuine phase transition. A phase transition is expected at zero quark mass, however[3]. Thus the pseudocritical temperature or crossover temperature is defined as the temperature of the peak of a suitable susceptibility, or the temperature of maximum change in an observable. Of course, the two definitions do not necessarily agree unless the chosen susceptibility is rigorously the derivative of the observable whose maximum slope is sought. Karsch and Laermann use the Polyakov loop susceptibility

$$\chi_L = N_s^3[\langle(\text{Re}P)^2\rangle - \langle\text{Re}P\rangle^2]. \qquad (1)$$

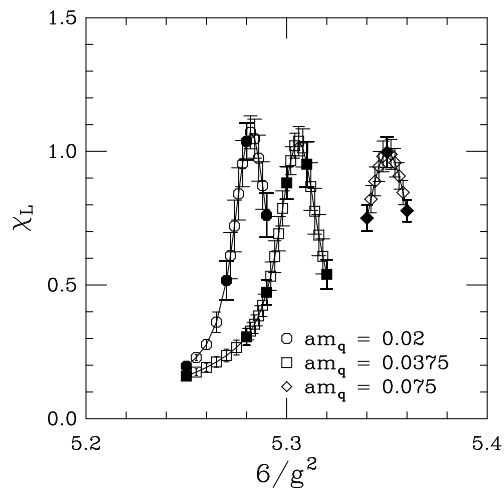

Figure 1. Polyakov loop susceptibility *vs* $6/g^2$ from Karsch and Laermann. Solid symbols are directly simulated. Open symbols are derived by reweighting.

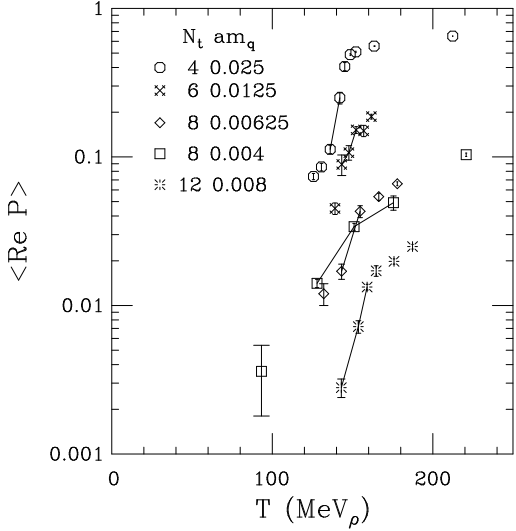

Figure 2. Polyakov loop *vs* temperature. The line segments indicate the range of uncertainty in locating the maximum slope.

Combined with a Ferrenberg-Swendsen reweighting analysis to interpolate between simulation points, the method permits a clean determination of the pseudocritical coupling, as shown in Fig. 1[4]. At larger lattice volumes, however, reweighting requires more closely spaced simulation points and becomes infeasible. The maximum slope method is used instead in Fig. 2 to locate the phase transition for a wide range of lattice sizes[4–7]. Two variables, namely $\langle ReP \rangle$ and $\langle \bar{\psi}\psi \rangle$ were examined to determine the maximum slope. This figure shows results only for the lowest available quark mass in each data set. Included are new results at $N_t = 8$ and preliminary results at $N_t = 12$[7], the latter corresponding to a lattice spacing of 0.1 fm. The temperature scale, based on the rho meson mass, is discussed in greater detail below. The temperatures thus determined are summarized in Fig. 3. There is an encouraging consistency in the results, which place the crossover temperature at approximately 140–160 MeV, based on the $\rho$ meson mass. It should be emphasized, however, that over the range of lattice parameters of this compilation, the nucleon to rho mass ratio is approximately 1.5, 20% above the physical value. Using the nucleon mass to set the temperature scale would therefore result in a 20% reduction in the crossover temperature. Physical values of this ratio have been reported in quenched calculations for both Wilson and staggered fermions[8], offering hope for future simulations with dynamical fermions. Furthermore, the mass ratio $m_\pi/m_\rho$ lies in the range 1/3 to 1/2, twice to three times the experimental value, because the quark mass is too high. Indeed, an extrapolation to zero quark mass was attempted in constructing Fig. 3. However, present indications suggest that lowering the quark mass does not substantially change the crossover temperature.

The temperature scale in Fig. 2 was constructed following Blum *et al* [6,9] and is based on the lattice rho meson mass. That is, from several zero temperature simulations[12] an interpolation fitting formula is constructed, giving $am_\rho(6/g^2, m_q a)$. For present purposes, we use the combination of cubic splines given by

$$
\begin{aligned}
am_\rho &= S_0(6/g^2) + S_1(6/g^2)am_q \\
&\quad + S_2(6/g^2)(am_q)^2 \qquad (2) \\
m_\pi/m_\rho &= (am_q)^{1/2}\hat{S}_0(6/g^2) \\
&\quad + \hat{S}_1(6/g^2)am_q \\
&\quad + \hat{S}_2(6/g^2)(am_q)^{3/2} \qquad (3)
\end{aligned}
$$

with knots at $6/g^2 = 5.3$, 5.5, and 5.7. Figure 4 shows the parameter range of thermodynamic simulations and companion zero temperature spectrum simulations used in this review, indicated by the plot symbol "s". As is usual, the thermodynamic simulations outpace the spectrum simulations. Thus for $6/g^2 > 5.7$ encountered in the $N_t = 12$ simulations, values higher than any available zero temperature spectrum study, an extrapolation becomes necessary. For this purpose I used an approximate tadpole-improved asymptotic scaling formula[10], based on high temperature plaquette measurements. A temperature in MeV is then inferred from $T = 770/[am_\rho(6/g^2, m_q a)N_t]$. This scale does not allow for variations in the rho mass with quark mass. It also inherits all the systematic errors of the zero temperature spectral calculations. Nonetheless, the rough figure 140–160 MeV for the crossover is already extremely useful for models of quark plasma formation.

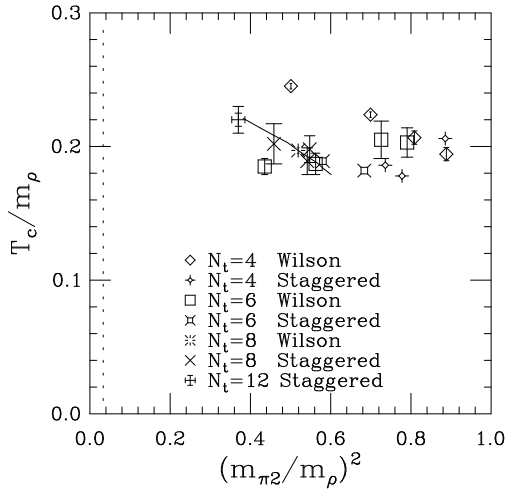

Figure 3. Temperature of the crossover in units of the rho meson mass with two light quarks *vs* the squared ratio of the pion to rho mass. For points from the staggered simulations, the mass of the second (non-Goldstone) pion is used. The curved line segment shows the error bar for the new $N_t = 8$ staggered point. The vertical dashed line indicates the physical mass ratio.

## 3. CONSTITUENT QUARK FREE ENERGY

The wide range of $N_t$ values available this year makes possible an amusing analysis of the Polyakov loop variable, which measures the change in the free energy of the thermal ensemble due to the introduction of a point spinless test

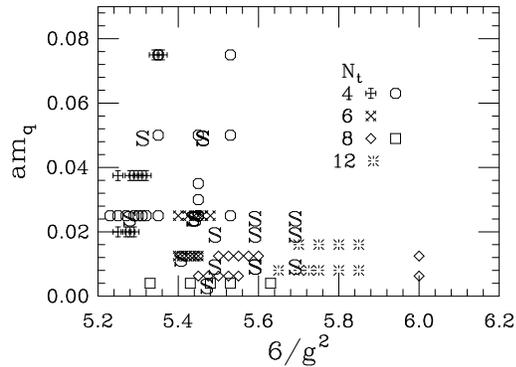

Figure 4. Thermodynamic simulation points (conventional plot symbols) and spectrum simulation points (letter s) for staggered fermion data used in this review.

quark. This free energy difference

$$f(T, m_q) = -T \log \langle \mathrm{Re} P/3 \rangle, \qquad (4)$$

a function of the temperature $T$ and light quark mass $m_q$, includes the lattice-regulated ultraviolet-divergent self-energy of the point source, proportional to the inverse lattice spacing $1/a$, and the free energy of the screening cloud of light antiquarks and quarks, which we might call with some risk of confusion the "constituent quark free energy". Computing the self energy to leading order in perturbation theory, we have

$$f(T, m_q) = 2\pi C_F \alpha_V \gamma/a + f_{\mathrm{cq}}(T, m_q). \qquad (5)$$

where $C_F = 4/3$ is the color Casimir factor for the triplet representation, $\alpha_V$ is the color fine structure constant for appropriate to heavy quark bound states at the same lattice scale, and

$$\gamma = \frac{1}{N_s^3} \sum_k \frac{1}{6 - 2\sum_\mu \cos(2\pi k_\mu/N_s) + (Ma)^2} \qquad (6)$$

is the dimensionless static lattice propagator for a Debye-screened electrostatic gluon field evaluated at zero separation. Why screening? Although the ultraviolet divergent contribution $1/a$ is uniquely determined in the limit $a \to 0$, an arbitrary infrared cutoff, here embodied in the Debye mass $M$, determines where the contribution from the point quark ends and the contribution from the screening cloud begins. Thus there is no unique definition of "constituent quark free energy". Instead, within the framework of any consistent definition, Eq (5) permits a separation of two contributions, one varying in a known way with the lattice scale $1/a$ and the other, unknown, but scale-invariant. Herein lies its predictive power.

For present purposes, using $N_t = 1/aT$, I chose a simpler approximate form

$$f_{cq}(T, m_q) = -T(\log\langle \mathrm{Re} P/3\rangle + cN_t) \qquad (7)$$

and adjusted the dimensionless constant $c$ by eye to achieve the rough scaling agreement shown in Fig. 5. For each data set only values for the lightest available quark mass are used. Although the quark mass values $m_q/T$ are not the same from one $N_t$ to the next in this figure, they are small ($m_q/T \leq 0.1$) and would be expected to contribute little (of the order 10 MeV) to the free energy. Thus one would expect only a small inconsistency from this variation. The best value for $c$ appears to be about 0.4, compared with values ranging from 0.26 to 0.35 expected from the lowest order perturbative self energy with a screening mass $M = 3.2T$. We note that at the crossover, the free energy drops by about the 300 MeV expected in a constituent quark model with deconfinement at high temperature.

## 4. CRITICAL BEHAVIOR

The phase structure of QCD also has important phenomenological consequences. Sketched in Fig. 6 is a proposed phase diagram with two flavors of light quarks and one strange quark (2 + 1 flavors), showing in what mass range a thermal phase transition is expected, and whether it is first or second order. Such a phase structure is motivated by an analysis of the corresponding sigma models in mean field theory, augmented by an analysis of quantum fluctuations[3,11]. Whether QCD conforms to this expectation remains to be established. Simulations with 2 + 1 flavors in the staggered and Wilson fermion schemes both support the existence of the first order region, but do not agree on its extent[1,14,15]. Staggered fermion simulations of the Columbia group found evidence for a first order signal for $(m_{u,d}, m_s) \approx (15, 15)$ MeV, but no signal for $(m_{u,d}, m_s) \approx (15, 30)$ MeV. A recent Wilson fermion simulation by the Tsukuba group found a first order signal with quark masses as large as $(m_{u,d}, m_s) \approx (150, 150)$ MeV and $(m_{u,d}, m_s) \approx (0, 400)$ MeV[1,15]. Is the discrepancy a strong coupling artifact? At strong coupling the fermion doublers in the Wilson scheme tend to increase the effective number of relatively light flavors. With more flavors the chiral phase transition is stronger. On the other hand in the staggered fermion scheme, the breaking of flavor symmetry in strong coupling may reduce the effective flavor number, thereby weakening the phase transition.

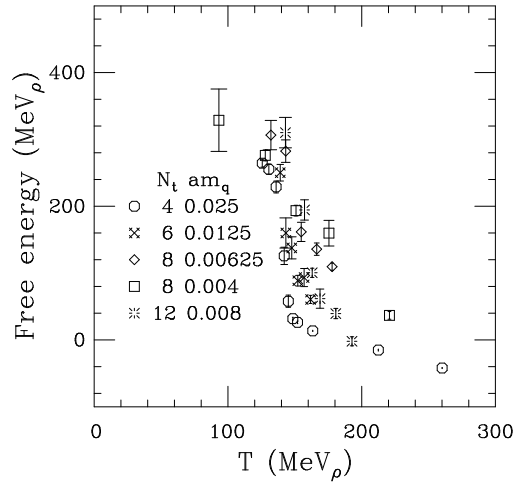

Figure 5. Constituent quark free energy as defined in text.

A recent analysis of the three-dimensional Gross-Neveu model by Kocić and Kogut questions the conventional wisdom that places QCD with its composite scalar mesons in the same universality class as sigma models with their elementary scalar mesons[16]. One would expect that the 3D Gross-Neveu model would exhibit 2D Ising universality. Instead, in a detailed simulation Kocić and Kogut found mean-field scaling. Now for QCD and the quark plasma there may be little phenomenological importance to such sub-tleties, but the question is of broad significance for understanding critical behavior in field theories. Furthermore, to take the continuum limit of lattice simulations, we must learn to do extrapolations to small quark mass and large volume. In the vicinity of the phase transition, a correct extrapolation requires the correct critical exponents. Thus an important test of this proposed phase structure is to determine whether QCD has the critical behavior expected from universality along the critical phase boundary.

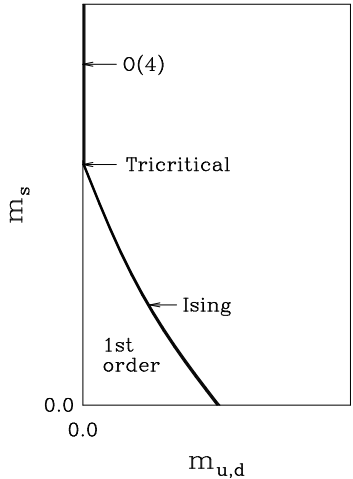

Figure 6. QCD phase diagram for 2 + 1 flavors as a function of a degenerate up and down quark mass $m_{u,d} = m_u = m_d$ and a strange quark mass $m_s$. The heavy line is the critical phase boundary.

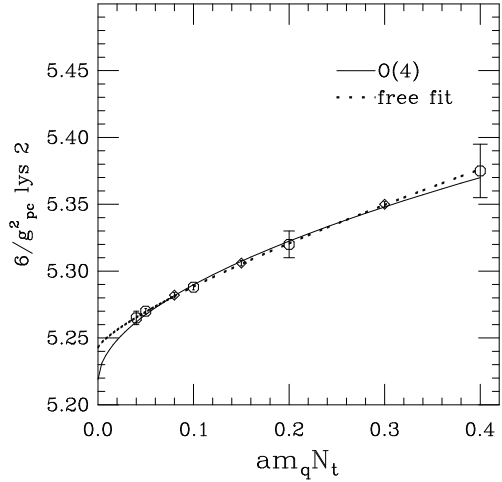

Figure 7. Crossover coupling *vs* quark mass for two flavors of staggered quarks from Karsch and Laermann. Solid line is a fit with O(4) critical exponents. Dashed line is a free fit. Diamonds are new results.

For the remainder of this section I will focus on the two flavor theory in the staggered fermion scheme, corresponding to the upper portion of the $m_s$ axis in Fig. 6. In this case O(4) universality is expected in the continuum limit. To be more precise, in the staggered fermion scheme, one expects O(2) critical behavior on coarse lattices where the flavor symmetry breaking of the staggered scheme is significant, and O(4) as the continuum limit is approached. Recent work by Karsch and Karsch and Laermann attempts a determination of some of the critical exponents of QCD[4,17]. They exploit the standard sigma model analogy between QCD and a magnetic system. In this analogy the quark mass plays the role of an external magnetic field and $\langle\bar\psi\psi\rangle$ plays the role of the magnetization.

The standard analysis of critical behavior begins with the assumed scaling of the critical contribution to the free energy in a magnetic system[18]. This contribution is singular and dominant at small field close to the zero field critical temperature $T_c(0)$. The scaling property, expressed in terms of the scaled temperature $t = [T - T_c(0)]/T_c(0)$ and magnetic field $h$, is

$$f_{\text{crit}}(t, h) = b^{-1} f_{\text{crit}}(b^{y_t} t, b^{y_h} h). \qquad (8)$$

From this scaling behavior one derives a scaling relation for the critical contribution to the magnetization $s = -\partial f_{\text{crit}}/\partial h$:

$$s(t, h) = h^{1/\delta} y(x) \qquad (9)$$

where $x = t h^{1/\beta\delta}$ and $y(x)$ is a scaling function (often called an equation of state in the statistical mechanics literature). Here $\beta$ and $\delta$ are standard critical exponents derived from $y_t$ and $y_h$. An important consequence of this result is that a crossover peak in the susceptibility $\chi_h = \partial s/\partial h$ moves along a curve of constant $x = x_{\rm pc}$ as $h$ and $t$ are varied. Thus if critical scaling holds, once the pseudocritical temperature is known at one $h$, it can be predicted at all $h$.

In QCD the quark mass plays the role of the magnetic field and $\langle\bar\psi\psi\rangle$, the magnetization. Specifically, Karsch suggests using $h = m_q/T = am_q N_t$ and $t = 6/g^2 - 6/g_c^2(0, N_t)$, where $g_c(0, N_t)$ is the critical gauge coupling at zero quark mass for a particular $N_t$[17]. Scaling then implies that

$$6/g_{\rm pc}^2(m_q a) = 6/g_c^2(0) + (m_q/T)^{1/\beta\delta}. \qquad (10)$$

Using this expression Karsch presented an analysis of the crossover for $N_t = 4$, 6, and 8 for two quark flavors[17]. He found encouraging agreement. The addition of new data this year at $N_t = 4$ permits a more refined analysis, shown in Figure 7[4]. Their best fit critical exponent $1/\beta\delta$ is $0.77 \pm 0.14$, in slight disagreement with the O(4) value 0.55(2), but consistent with the O(2) value 0.60(1) and the mean field value 0.67.

Karsch and Laermann also introduced a new cumulant

$$\Delta = \frac{\partial \ln\langle\bar\psi\psi(6/g^2, m_q)\rangle}{\partial \ln m_q} = \frac{1}{\delta} - \frac{x y'(x)}{\beta\delta\, y(x)} \qquad (11)$$

that evaluates the critical exponent $\Delta = 1/\delta$ at $x = t = 0$. They obtain $0.21 < 1/\delta < 0.26$ consistent with the O(4) value 0.208(2) and O(2) value 0.2080(3), and somewhat inconsistent with the mean field value 0.33. Thus the Kocić-Kogut scenario cannot be decisively excluded.

To carry these results further, I test the scaling relation (9) in simulations with two flavors of staggered fermions over the wide range of currently available $N_t$. For this purpose I use slightly different variables to permit comparison among different $N_t$ and to avoid quantities with anomalous dimensions, namely,

$$h = m_\pi^2(m_q, T=0)/m_\rho^2(m, T=0) \qquad (12)$$
$$t = [T - T_c(0)]/T_c(0) \qquad (13)$$
$$s = h^{-1} m\langle\bar\psi\psi(m_q, T)\rangle/T^4 \qquad (14)$$

The scaling relation (9) then gives a universal function

$$y(x) = h^{-1-1/\delta} m_q \langle\bar\psi\psi(m,T)\rangle/T^4 \qquad (15)$$

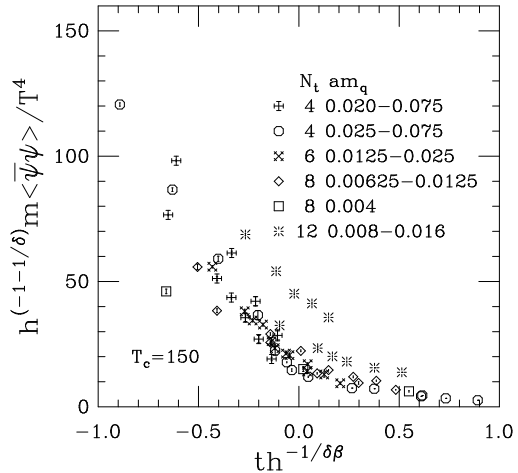

Figure 8. Scaled $\langle\bar\psi\psi\rangle$ vs scaled temperature with O(4) critical exponents.

with $x = t h^{1/\beta\delta}$. The extra factor $h^{-1}$ is needed to compensate the quark mass factor $m_q$. Let us apply this analysis to data for $\langle\bar\psi\psi\rangle$ from several groups[5]. Setting the critical exponents $\delta$ and $\beta$ to their O(4) values and adjusting the sole remaining free parameter $T_c(0)$ in 10 MeV increments to get the best agreement I get the result shown in Fig. 8. With the exception of the $N_t = 12$ data, the scaling agreement is rather good. At this level it is not possible to distinguish O(4) from O(2) and mean field critical behavior. Setting $T_c(0)$ to 140 MeV or 170 MeV worsens the agreement noticeably, but 160 MeV gives comparable consistency. Obviously a host of systematic errors, including finite volume effects and deviations from continuum scaling enter the analysis, so refinements are certainly needed before the method can serve as a definitive test of critical behavior. The most glaring inconsistency in this figure comes from the preliminary $N_t = 12$ data. Increasing $T_c(0)$ for only this data set to 160 MeV and plotting it with other data computed for $T_c(0) = 150$ MeV brings the $N_t = 12$ data at the lower quark mass into good agreement. Thus the discrepancy could be caused either by a gradual upward shift in the crossover temperature as the lattice spacing is decreased, or by an erroneous extrapolation of the hadron spectrum above $6/g^2 = 5.7$, or by a failure of the scaling hypothesis over this parameter range.

## 5. THERMODYNAMICS WITH WILSON FERMIONS

In order to claim we can infer features of continuum QCD from numerical simulations it is essential that we demonstrate that our answers are independent of the fermion scheme. Unfortunately, thermodynamic simulations with Wilson fermions are not sufficiently developed at present to make a confident comparison with the staggered scheme. The fundamental difficulty is that we are dealing with a phase transition associated with the restoration of a spontaneously broken chiral symmetry, but the Wilson scheme builds in an explicit breaking of this symmetry, which goes away only in the continuum limit.

It is popular to define a chiral line $\kappa = \kappa_c(6/g^2)$ where a number of characteristics of chiral symmetry (vanishing zero temperature pion mass, vanishing current quark mass) hold, at least with some consistency. Thus we are interested in studying the crossover or phase transition in a region close to the chiral line. The Tsukuba group found that to do so requires either working at very small values of $6/g^2$, usually considered to be in the strong coupling regime, or working at much higher $N_t$ than is feasible with present resources[19]. The former choice risks encountering lattice artifacts, and the latter is expensive. Indeed, evidence for lattice artifacts in two flavor simulations at $N_t = 6$ was recently reported by the MILC collaboration, which found a first order, possibly bulk, phase transition in simulations at $\kappa = 0.17$, 0.18, and 0.19 in close proximity to the thermal crossover[20]. The Tsukuba group has continued its pioneering work at small $6/g^2$. For an authoritative review, see Iwasaki's talk[1].

At this conference results of the first simulation at $N_t = 8$ in the Wilson scheme were reported by the MILC collaboration[21]. The simulation was done at $6/g^2 = 5.3$ over a range of $\kappa$ up to $\kappa_c \approx 0.168$, allowing a comparison with previous results at $N_t = 4$ and 6. The thermal crossover, now shifted to $\kappa_t \approx 0.167$, shows no ev-idence for the metastability seen at slightly larger $\kappa$ at $N_t = 6$. As illustrated for $\langle \bar\psi\psi \rangle$ in Fig. 9, as $N_t$ is increased, bulk quantities appear to follow an envelope established by the zero temperature theory, breaking away at the crossover. Moreover $\langle \bar\psi\psi \rangle$ appears to be decreasing immediately prior to the crossover, suggesting progress toward a low quark mass. For details, please see Toussaint[21]. Thus, fortunately, the disease seen at slightly larger $\kappa$ in the $N_t = 6$ simulations appears not to be spreading to lower $\kappa$ as $N_t$ increases.

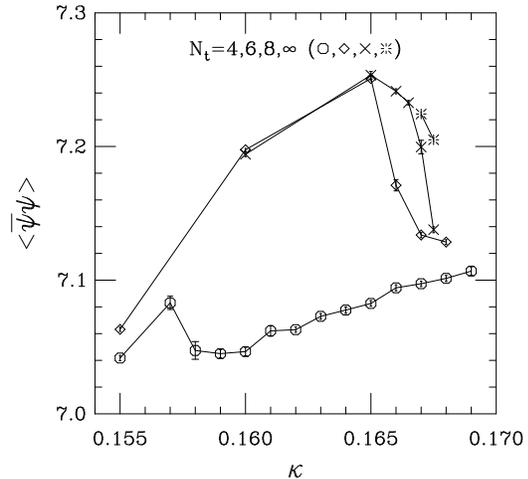

Figure 9. Chiral condensate $vs$ $\kappa$ for two flavors of Wilson fermions at $6/g^2 = 5.30$.

It is clear that with present methods, Wilson fermion thermodynamics are far more costly than staggered fermion thermodynamics, requiring far larger lattices to suppress strong-coupling artifacts. Further progress with the Wilson scheme is likely to require working with an improved action. Indeed the Tsukuba group has adopted one such improvement and they report some preliminary results at this conference[1,22].

## 6. PURE YANG MILLS WITH MIXED ACTION

This year Gavai, Grady, and Mathur reported results for simulations of SU(2) Yang-Mills theory

with a mixed fundamental/adjoint action

$$S = \beta_f \sum_P [1 - \frac{1}{N} \mathrm{Re} \mathrm{Tr}_f U_P]$$
$$+ \beta_a \sum_P [1 - \frac{1}{N^2} \mathrm{Tr}_f U_P^\dagger \mathrm{Tr}_f U_P]. \quad (16)$$

where $N = 2$ for SU(2), $\beta_f$ and $\beta_a$ are the fundamental and adjoint gauge couplings, the gauge link matrices $U$ are in the fundamental representation, and $\mathrm{Tr}_f$ is the trace in the fundamental representation[23]. They have reopened the question of the interplay between the well-known bulk transition[25] and the thermal phase transition.

Continuum universality requires that as $N_t$ is increased, the entire thermal phase boundary separating the confined and deconfined phases must shift toward weak coupling. If instead one end remains anchored to a bulk phase boundary, the location of which by definition is asymptotically constant as $N_t \to \infty$, it would be possible to approach the continuum limit at zero temperature along some directions in the $\beta_a$—$\beta_f$ plane, all the while remaining in the *deconfined* phase, and to approach the continuum limit at zero temperature along other directions in the *confined* phase. Gavai, Grady and Mathur found that at $N_t = 4$ the second order thermal phase boundary in SU(2) connects smoothly to what had been hitherto called the first order bulk phase boundary. If universality were to survive, the $N_t = 6$ phase boundary would have to show a shift toward weak coupling. Indeed, in a more recent remarkable work Mathur and Gavai found that at $N_t = 6$, the strangely coalesced phase boundaries appear to have shifted together toward weak coupling slightly[26], suggesting, perhaps, that the first order segment had been mislabeled as a bulk phase transition.

At this conference Heller presented results of recent simulations in SU(3) Yang-Mills theory, for which an apparently different and perhaps less surprising picture is emerging[27]. Results are summarized in Fig. 10. For $N_t = 4$ the thermal and bulk phase transitions are found to coalesce in SU(3) as in SU(2). But for $N_t = 6$ and 8 the thermal phase boundary clearly breaks away from the terminal segment of the bulk line, joining it above the bulk endpoint. As $N_t$ is increased, the junction moves to higher $\beta_a$, so that the entire thermal line shifts toward weak coupling in a manner consistent with universality. It would be surprising if the SU(2) result were to remain so dramatically different at still higher $N_t$.

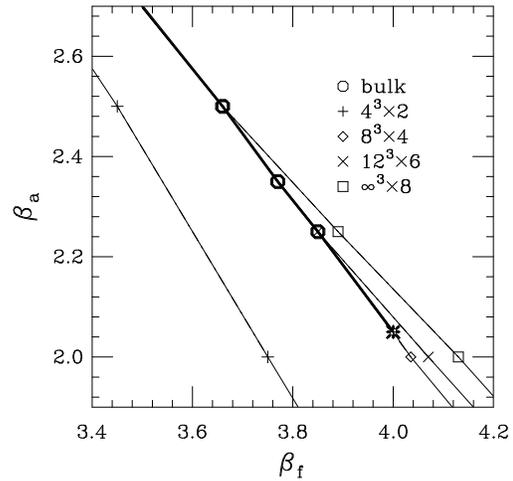

Figure 10. Phase diagram for pure Yang-Mills theory with a mixed fundamental-adjoint action. The bulk phase boundary, applicable to $N_t = 4$, 6, and 8, is plotted as a darker line. The burst marks the bulk endpoint. The $N_t = 8$ points were found from an infinite volume extrapolation.

The signature for a separation between the bulk and thermal phase transitions in mixed action Yang-Mills theory is similar to what was found with Wilson fermions at large $\kappa$ in the $N_t = 6$ simulations, namely, with increasing $\beta_f$, a first order phase transition is first encountered, signaled by a discontinuity in the plaquette, but without an appropriate jump in the Polyakov loop variable. At slightly higher $\beta_f$ the Polyakov loop variable then rises, signaling a thermal phase transition or crossover. Now the Wilson fermion determinant surely induces an adjoint plaquette, which might become strong at large $\kappa$. Thus there may be more than coincidence relating the two phenomena. However, at this conference Rummukainen reported simulations measuring the strength of induced fundamental and adjoint couplings using a microcanonical demon method. He found a disappointingly small induced adjoint action, raising doubts about such an explanation for the peculiar behavior of Wilson thermodynamics at large $\kappa$[28].

## 7. EQUATION OF STATE WITH NONPERTURBATIVE METHODS

The equation of state for hadronic matter, giving the energy density $\epsilon(T, m_q)$ and pressure $p(T, m_q)$ as a function of temperature and quark mass is another quantity of phenomenological importance. The earliest determinations of these quantities used the basic thermodynamic identities

$$\epsilon V = \frac{\partial F}{\partial (1/T)} \quad (17)$$

$$p/T = \frac{\partial F}{\partial V} \quad (18)$$

In a lattice simulation each such derivative of the free energy involves a separate variation of the spatial and temporal lattice constants $a_s$ and $a_t$, requiring a concomitant renormalization of the gauge coupling. Some years ago Karsch determined the perturbative asymptotic variation of the gauge coupling with respect to the anisotropy parameter $\xi = a_t/a_s$[29]. Unfortunately, present simulations are not in the perturbative asymptotic region. Although in principal the Karsch coefficients could be determined nonperturbatively from lattice simulations, this has not yet been done successfully[9,30].

Fortunately there is a different nonperturbative route to $\epsilon$ and $p$. The so-called interaction measure

$$I = \epsilon - 3p \quad (19)$$

is more easily determined, since it involves an isotropic variation of the lattice constant, requiring only the usual nonperturbative renormalization of the lattice coupling. The pressure, on the other hand, can be determined separately from the free energy $f = -p$ by integrating either of two relations[31]

$$\langle \Box \rangle = \frac{1}{2V} \frac{\partial F}{\partial (6/g^2)} \quad (20)$$

$$\langle \bar{\psi}\psi \rangle = \frac{1}{V} \frac{\partial F}{\partial m_q}. \quad (21)$$

A vacuum subtraction is performed to give the pressure relative to the pressure of the nonperturbative vacuum:

$$\frac{p}{T^4} = 2N_t^4 \int_{\text{cold}}^{6/g^2} d(6/g'^2) \left[ \langle \Box(6/g'^2, am_q) \rangle_{N_t} \right.$$
$$\left. - \langle \Box(6/g'^2, am_q) \rangle_{\text{sym}} \right] \quad (22)$$

$$\frac{p}{T^4} = N_t^4 \int_{\text{cold}}^{am_q} d(am_q') \left[ \langle \bar{\psi}\psi(6/g^2, am_q') \rangle_{N_t} \right.$$
$$\left. - \langle \bar{\psi}\psi(6/g^2, am_q') \rangle_{\text{sym}} \right] \quad (23)$$

Of course the latter equation may be used only when dynamical quarks are present. Together with a determination of the interaction measure, this result can then be used to determine the energy density.

This year new nonperturbative determinations were reported for the energy density and pressure in SU(2) and SU(3) Yang-Mills theory[2,32] and in SU(3) with two flavors of dynamical staggered fermions[6,33]. In the latter case, although there is no evidence for a bona fide phase transition at nonzero quark mass, nonetheless there is a steep rise in the energy density at the temperature associated with the largest slope in the Polyakov loop and $\langle \bar{\psi}\psi \rangle$, as seen in Fig. 11. In a cooling quark plasma such a strong crossover could cause a momentary slowing in the expansion of the plasma as the quarks and gluons reorganized themselves into more compact hadrons.

## 8. MAGNETIC MONOPOLES AND DECONFINEMENT

A study of the thermal behavior of QCD may provide insight into the mechanism of confinement. An old 't Hooft—Mandelstam model characterizes the confining QCD vacuum as a dual superconductor, with an electric Meissner effect and a condensate of color magnetic monopoles[34]. Some years ago Schierholz *et al* and Kronfeld *et al* explored the association between confinement and the presence of monopole currents in Yang-Mills theories[35]. Interest has revived recently[36,37].

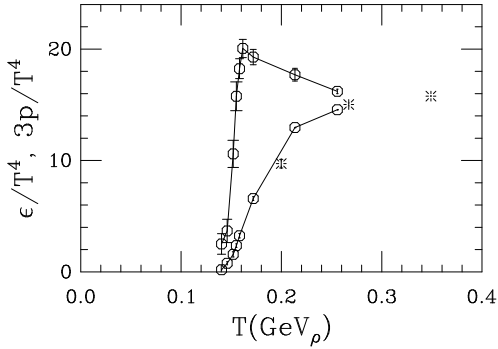

Figure 11. Energy density (upper curve) and three times the pressure (lower curve) vs temperature (scale based on the rho meson mass) for two light quark flavors ($m_q/T = 0.1$) in the staggered fermion scheme from Blum, Gottlieb, Kärkkäinen, and Toussaint. The bursts give the pressure extrapolated to zero quark mass

To identify monopole currents in a nonabelian gauge theory it is necessary to carry out a U(1) projection of the gauge links. This is done by first fixing a suitable gauge. Popular choices include maximal Abelian gauge and a variety of "unitary" gauges, one of which involves diagonalizing the product of gauge links forming the Polyakov loop. A "monopole" contribution is then extracted from the resulting U(1) gauge field following the procedure of DeGrand and Toussaint[38].

In the new approach the string tension and other confinement features are computed using only the monopole contribution to the U(1) field. Good agreement is found with the full string tension computed in the conventional way. In the past year the Kanazawa group has also calculated the projected U(1)-monopole Polyakov-loop expectation value in both SU(2) and SU(3) Yang-Mills theory[39]. The behavior of the monopole-projected Polyakov loop variable imitates the behavior of the conventional Polyakov loop variable. The similarity is seen in a variety of U(1) projection schemes.

Although the results are promising, further work is needed (1) to demonstrate that the suitably defined abelian monopole currents survive the continuum limit[40] and (2) to find a suitable order parameter for monopole condensation[41].

## 9. INSTANTON MOLECULES AND THE CHIRAL PHASE TRANSITION

The Stony Brook instanton program classifies the dominant gauge field configurations in the thermal ensemble in terms of their instanton content and seeks to understand the principal features of QCD in terms of them. In recent work Ilgenfritz and Shuryak and Schäfer, Shuryak, and Verbaarschot argue that in full QCD, as the chiral phase transition is approached, the light fermion determinant induces an attractive interaction between instantons and anti-instantons[42]. The resulting molecules are predicted to predominate over solitary instantons and anti-instantons. It is argued that the strong correlation drives the chiral phase transition. Schäfer, Shuryak, and Verbaarschot compute hadronic screening masses in the model and find a spectrum in qualitative agreement with results from lattice simulations. It would be interesting to test their proposals further in lattice simulations.

## ACKNOWLEDGMENTS

I am indebted to my colleagues for their essential help in preparing this review. In particular, I thank Doug Toussaint, Bob Sugar, Urs Heller, Leo Kärkkäinen, Kari Rummukainen, Tom Blum, Steve Gottlieb, and Krishna Rajagopal. I am also grateful to Frithjof Karsch, Edwin Laermann, and Robert Mawhinney for providing assistance with their data. I would like to thank the Institute for Theoretical Physics, University of California at Santa Barbara, where part of this work was carried out. This work was supported in part by the U.S. National Science Foundation under grant NSF-PHY9309458 and by the U.S. Department of Energy under contract DE-FG03-93ER25186.